\documentclass[twocolumn,preprintnumbers,aps,amsmath,amssymb]{revtex4}

\usepackage{amssymb,amsfonts,amsmath}
\usepackage{graphicx}
\usepackage{dcolumn}
\usepackage{bm}

\newcommand{\ave}[1]{\langle #1\rangle}
\newcommand{\eq}[1]{Eq. (\ref{#1})}
\def\reactionrates#1#2{\mathrel{\mathop{\rightleftharpoons}\limits^{#1}_{#2}}}
\begin{document}
\title[Amplification and detection of single molecule fluctuation ...]
{Amplification and detection of single molecule conformational fluctuation through a  protein interaction network with bimodal distributions}

\author{Zhanghan Wu$^{1}$, Vlad Elgart$^{1}$
, Hong Qian$^{2}$, Jianhua Xing$^{1}$} \email{jxing@vt.edu} \affiliation{$^{1}$
Department of Biological Sciences, Virginia Polytechnic Institute and State University, Blacksburg, VA}
\affiliation{$^{2}$ Department of Applied Mathematics, University of Washington, Seattle, WA}


\begin{abstract}
  A protein undergoes conformational dynamics with multiple time scales, which results in fluctuating enzyme activities. Recent studies in single molecule enzymology have observe this ``age-old'' dynamic disorder phenomenon directly. However, the single molecule technique has its limitation. To be able to observe this molecular effect with real biochemical functions {\it in situ}, we propose to couple the fluctuations in enzymatic activity to noise propagations in small protein interaction networks such as zeroth order ultra-sensitive phosphorylation-dephosphorylation cycle.  We showed that enzyme fluctuations could indeed be amplified by orders of magnitude into fluctuations in the level of substrate phosphorylation | a quantity widely interested in cellular biology.  Enzyme conformational fluctuations sufficiently slower than the catalytic reaction turn over rate result in a bimodal concentration distribution of the phosphorylated substrate.  In return, this network amplified single enzyme fluctuation can be used as a novel biochemical ``reporter'' for measuring single enzyme conformational fluctuation rates.
\end{abstract}

\maketitle

\section{Introduction}
The dynamics of an enzyme is usually characterized by a rate constant describing its catalytic capacity, which is a standard practice on studying dynamics of enzymes and enzyme-involved networks.
Recent advances in single molecule techniques allow examining enzyme activities at single molecule levels  \cite{LuXun1998,Chen2003, Tan2003,Lu2005, Liu2006, English2006, Rissin2008, Wang2008, chen2009, Chen2008, Roeffaers2007, Xu2009, MinLuo2005}.  It is found  that the rate "constant" of an enzyme is in general a broad distribution. Physically, it is because the enzyme conformation is under constant fluctuation at varying time scales \cite{MinLuo2005, XingKim2006}. Single molecule techniques can measure the instant rate constants at a given conformation. The single molecule results are consistent with extensive early biochemistry and biophysics studies. Biochemists have long noticed that protein conformational fluctuations (which can be in the time scale from subsecond to minutes and even hours) can be comparable and even slower than the corresponding chemical reactions (usually in the range of subsecond) \cite{Frieden1979}. Slow conformational motions result in hysteretic response of enzymes to concentration changes of regulatory molecules, and cooperative dependence on substrate concentrations \cite{Frieden1979,Frieden1970,Ricard1987,Ainslie1972}. In physical chemistry, the term dynamic disorder is used for the phenomenon that the rate of a process may be stochastically time-dependent \cite{Zwanzig1990}. Extensive experimental and theoretical studies exist since the pioneering work of Frauenfelder and coworkers \cite{Austin1975}. Allosteric enzymes can be viewed as another class of examples. According to the classical Monod-Wyman-Changeux model and recent population shift model, an allosteric enzyme coexists in more than one conformation \cite{Monod1965,Kern2003}. Recent experiments also show that the conformational transition of an allosteric enzyme happens in micro- to millisecond time scale or longer \cite{Volkman2001}. Xing proposed that in general internal conformational change should be considered on describing enzymatic reactions, and it may have possible implication on allosteric regulation mechanism \cite{Xing2007}. Wei {\it et. al} also suggested a similar formalism for describing enzymatic reactions \cite{Min2008}. Current single molecule enzymology studies focus on metabolic enzymes. It remains an important unanswered question if dynamic disorder is a general phenomenon for enzymes, e.g., the enzymes involved in signal transduction. While the technique reveals important information, the single molecule approach also has strict requirement on the system to achieve single molecule sensitivity, which limits its usage for quick and large scale scanning of enzymes.

In this work we discuss an idea of coupling molecular enzymatic conformational fluctuations to the dynamics of small protein interaction networks. Specifically we will examine a phosphorylation/dephosphorylation cycle (PdPC) \cite{Goldbeter1981}. Our analysis will be applicable to other mathematically equivalent systems, such as GTP-associated cycle, or more general a system involving two enzymes/enzyme complexes with opposing functions on a substrate. As an example for the latter, the system can be an enzymatic reaction consuming ATP hydrolysis ({\it e.g.}, a protein motor) coupled to a ATP regeneration system--in this case ATP is the substrate. The PdPC is a basic functional module for a wide variety of cellular communications and control processes. The substrate molecules can exist in the phosphorylated and dephosphorylated form, which are catalyzed by kinase and phosphatase respectively at the expense of ATP hydrolysis. The percentage of the phophorylated substrate form depends on the ratio of kinase and phosphatase activities in a switch like manner called ultra-sensitivity. Through the PdPC, slow conformational (and thus enzymatic activity) fluctuations at the single molecule level can be amplified to fluctuations of substrate phosphorylation forms by several orders of magnitude, and make it easier to detect.
The coupling between molecular fluctuations and network fluctuations itself is an interesting biological problem. Recent studies revealed that the intrinsic/extrinsic noise, when it is introduced into the biological system, has significant influences on the behavior and sensitivity of the entire network \cite{Swain2002,Levine2007,Samoilov2005,Samoilov2006,Lepzelter2007,Thattai2001,Barkai1997, Miller2008}. Unlike the noise sources studies previously, the internal noise due to dynamic disorder shows broad time scale distributions. Its effect on network-level dynamics is not well studied \cite{Xing2008}.  We will show that a bimodal distribution of the PdPC substrate form can arise due to dynamic disorder, which may have profound biological consequences.

\section{The model}
A PdPC is shown in  $\ref{Figure1}$a. X and X* are the unphosphorylated and phosphorylated forms of the substrate, respectively. We assume A is the phosphatase obeying normal Michaelis-Menten kinetics. The kinase E, on the other hand, can assume two conformations with different catalytic capacity. In general an enzyme can assume many different conformations. The two state model here can be viewed as coarse-graining.
The set of reactions describing the system dynamics are listed below:
\begin{eqnarray*}
X+E_1\reactionrates{k_{1f}}{k_{1b}} XE_1\reactionrates{k_{+f}}{k_{+b}}   X^* + E_1,\\
X+E_2\reactionrates{k_{1f'}}{k_{1b'}}  XE_2\reactionrates{k_{+f'}}{k_{+b'}}   X^* + E_2,\\
X^*+A\reactionrates{k_{2f}}{k_{2b}} X^*A \reactionrates{k_{-f}}{k_{-b}}  X + A,\\
E_1\reactionrates\alpha\beta E_2, E_1 X\reactionrates{\alpha'}{\beta'} E_2X.
\end{eqnarray*}
To ensure proper detailed balance constraint, each pair of forward and backward reaction constants are related by the relation $\Delta\mu^0 = -k_BT\ln(k_f/k_b)$, with $\Delta\mu^0$ the standard chemical potential difference between the product(s) and the reactant(s), $k_B$ the Boltzmann constant, $T$ the temperature. Exceptions are the three chemical reaction steps, which we assume couple to ATP hydrolysis, and thus extra terms related to ATP hydrolysis free energy $\Delta\mu_{ATP}$ are added. we assume $\Delta\mu^0 +\Delta\mu_{ATP}/2= -k_BT\ln(k_f/k_b)$ for each of these reactions, so one ATP molecule is consumed after one cycle.  In general the conformer conversion rates $\alpha,\beta,\alpha',\beta'$ are different. For simplicity in this work, we choose $\alpha=\beta=\alpha'=\beta'$ unless specified otherwise.

Let's define the response curve of the system to be the steady-state percentage of X* as a function of the catalytic reactivity ratio between the kinase and the phosphatase ($\overline{\theta}\equiv (p_1k_+ + p_2k_+')N_{E_t}/(k_-N_{A_t})$, where $p_i$ is the probability for the kinase to be in conformer $i$),  and $N_{E_t}$ and $N_{A_t}$ are total numbers of kinase and phosphatase molecules. At bulk concentration, for both fast and slow kinase conformational change, the response curve shows usual sigmoidal but monotonic dependence (see  $\ref{Figure1}$b) \cite{Goldbeter1981}. Here we only consider the zero-th order regime where the total substrate concentration is much higher than that of the enzymes.

However, the situation is different for a system with small number of molecules. For simplicity let's focus on the case with one kinase molecule. Physically, suppose that the average reactivity ratio $\overline{\theta}\sim 1$ (not necessarily exactly at 1), but the corresponding $\theta_1\equiv  k_+ /(N_{A_t}k_-)>1$ and $\theta_2\equiv k_+' /(N_{A_t}k_-)<1$. Consequently, the substrate conversion reactions are subject to fluctuating enzyme activities, an manifestation of the molecular level dynamic disorder. Because of the ultra-sensitive nature of the PdPC, small enzyme activity fluctuation (in the vicinity of $\theta = 1$) can be amplified into large fluctuations of substrate forms (in the branches of high or low numbers of X*).  The relevant time scales in the system are the average dwelling time of the kinase at the new conformation $\tau_K$, and the average time required for the system to relax to a new steady-state substrate distribution once the kinase switches its conformation $\tau_S$. The former is related to the conformer conversion rates. The latter is determined by the enzymatic reaction dynamics as well as the number of substrate molecules. If the kinase conformational switch is sufficiently slow ($\tau_K>>\tau_S$), so that on average for the time the kinase dwelling on each conformation, the substrate can establish the steady state corresponding to $\theta_i$, which is peaked at either high or low $N_{X^*}$. Then the overall steady-state substrate distribution is  a bimodal distribution, which is roughly a direct sum of these two single peaked distributions. This situation resembles the static limit of molecular disorders \cite{Zwanzig1990}. Increasing conformer switching rates tends to accumulate population between the two peaks, and eventually results in a single-peaked distribution ($\tau_K<\tau_S$). A critical value of $\alpha$ (or $\beta$) exists where $\tau_K\sim \tau_S$, and one peak of the distribution disappears. There are two sets of ($\tau_K,\tau_S)$ corresponding to the transition from conformer $1$ to $2$, and vice versa. In principle, in the slow enzyme conform conversion regime where the substrate shows non-unimodal distribution ($\tau_K>\tau_S$), one can extract molecular information of the enzyme fluctuations from the greatly amplified substrate fluctuations. This is the basic idea of this work.

\section{Numerical studies}

\begin{figure}
 \begin{center}
    \includegraphics*[width=3.25in]{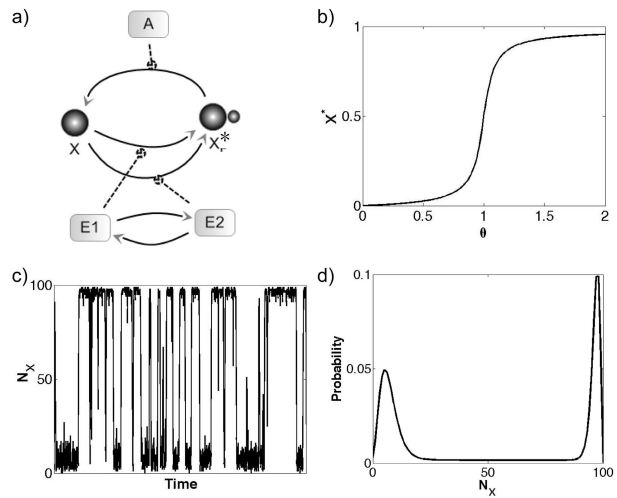}
    \caption{A PdPC with a single kinase enzyme shows bimodal distribution of the substrate.
(a) Illustration of the PdPC. The kinase molecule has two slowly converting conformers with different enzyme activity.
(b) The bulk response curve shows sigmoidal and monotonic zeroth-order ultrasensitivity. These results are obtained by solving the rate equations with the parameters given in Table 1.
(c) A trajectory of the number of X with a single two state kinase. The total substrate number $N_{Xt}=100$, conversion rate constant between two kinase states $\alpha=0.001$, and other parameters are shown in Table 1.
(d) The distribution of X corresponding to (c) shows bimodal distribution.
}
    \label{Figure1}
 \end{center}
\end{figure}

To test the idea, we performed stochastic simulations with the Gillespie algorithm \cite{Gillespie1977} using the parameters listed in Table 1 and Appendix A.  \ref{Figure1}c shows that with a single slowly converting two-state kinase, the number of X* jumps between high and low values, and shows bimodal distribution (see   \ref{Figure1}d).   \ref{Figure2} gives systematic studies on this phenomenon. There exist two critical values of $\alpha$, $\alpha_1,\alpha_2$. With $\alpha<(\alpha_1,\alpha_2)$, the substrate distribution has two well-separated peaks (  $\ref{Figure2}$a). On increasing $\alpha$, the two original peaks diminish gradually while the the region between the two peaks accumulates population to form a new peak (  $\ref{Figure2}$b). The two original peaks disappear at $\alpha=\alpha_1$ and $\alpha=\alpha_2$ respectively, and eventually the distribution becomes single-peaked (  $\ref{Figure2}$c-e).   \ref{Figure2}f summarizes the above process using the distance between peaks. The results divide into three regions. The point of transition between the left and the middle regions indicates disappearance of the left peak (corresponding to  \ref{Figure2}c). That between the middle and the right regions indicates disappearance of the right peak (corresponding to   \ref{Figure2}d).

  \ref{Figure2}g shows that the critical value of $\alpha$ decreases with the total substrate number $N_{Xt}$. An increased number of $N_{Xt}$ gives a larger $\tau_S$, which requires a  larger $\tau_K$ (slower conformation conversion rate) in order to generate a multi-peaked distribution.
  \ref{Figure2}g also compares theoretical (see below) and simulated critical $\alpha$ values at different values of $N_{X_t}$.
The plot shows that the simulation results agree reasonably with the theoretical predictions, although the simulated critical $\alpha$ is slightly smaller than the theoretical values, which means that the peak disappears earlier on increasing fluctuation rate $\alpha$.
The discrepancy of the two could be due to stochasticity of enzymatic reactions, which is fully accounted for in the simulations, but neglected in the theoretical treatment. The broader distribution leads to an earlier disappearance of peaks. This argument is supported by the fact that the difference between theoretical and simulated critical $\alpha$ is getting closer when the substrate number becomes larger, so fluctuations due to enzymatic reactions are further suppressed.

In the above discussions, we focus on a system with a single copy of the two-state kinase molecule.  Appendix B shows that a  simulation result with multi-state model gives similar behaviors.  \ref{Nenzyme}a shows that with multiple copies of enzymes, the substrate distribution of a PdPC can show similar transition from bimodal (or multi-modal in some cases) to unimodal behaviors, but the critical values of $\alpha$ are smaller (corresponding to slower conformational change) than those for the single kinase case . In these calculations, we scale the system proportionally to keep all the concentrations constant.

Possible biological significances of the bimodal distributions will be discussed below. Here we propose that additionally one can use the phenomenon to extract single molecule fluctuation information, especially the conformer conversion rates. Conventionally the information is obtained through single molecule experiments \cite{English2006, MinLuo2005}. For simplicity here we focus on the single enzyme case only. Suppose that an enzyme fluctuates slowly between different conformers, and one can couple a single molecule enzyme with a PdPC (or a similar system) with fast enzymatic kinetics. Then the conformational fluctuation dynamics at the single molecule level will be amplified to the substrate form fluctuations by orders of magnitude.   \ref{Dwell} gives the result of such an experiment simulated by computer. The trajectory clearly shows two states. To estimate the time  the system dwelling on each state, we define the starting and ending dwelling time as the first time the number of substrate molecules in the X* form reaches the peak value of $N_{X^*}$ distribution corresponding to that state in the forward and backward direction of the trajectory (see   \ref{Dwell}a). The above algorithm of finding the dwelling time may miss those with very short dwelling time so the substrate may not have enough time to reach the peak value, as seen in the trajectory. Nevertheless, the obtained dwelling time distributions are well fitted by exponential functions. The exponents give the values of $\alpha$ and $\beta$, in this case $\sim 0.8\times 10^{-3}$ for both of them, which are good estimations of the true value $10^{-3}$.

\section{Theoretical analysis}

\subsection{Analytical estimate of the critical points}
\begin{figure}
 \begin{center}
     \includegraphics*[width=3.25in]{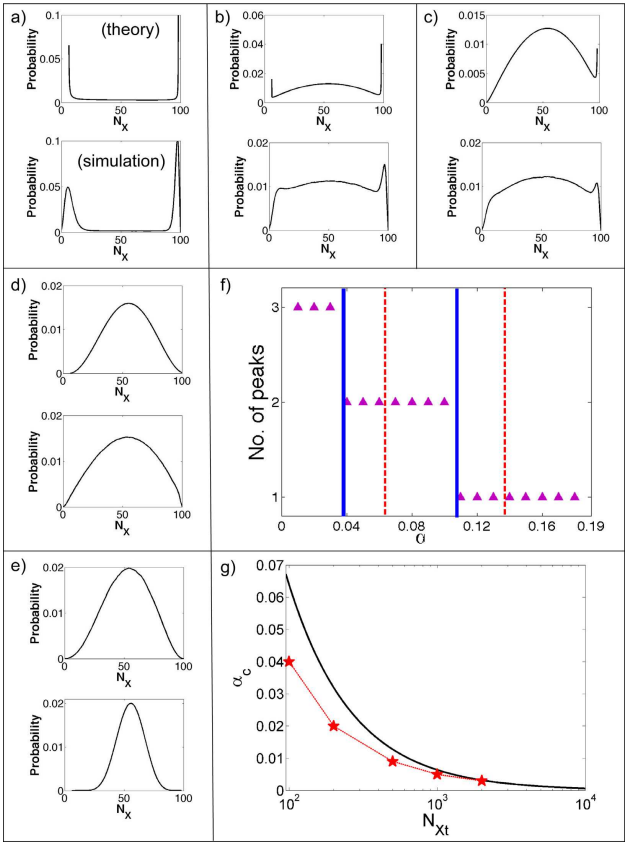}
     \caption{Dependence of substrate distribution of a PdPC with a single two-state kinase enzyme on the kinase conformer conversion rates. Total substrate molecule $N_{X_t}=100$ unless specified otherwise.
Simulation (s): lower, Theory (t): upper. For the parameters chosen, the theoretically predicted two critical points where the two peaks (singular points in the theory) disappear (\eq{f_critical}) are $\alpha_1 = 0.064, \alpha_2 = 0.137$.
(a) $\alpha_{t}= \alpha_{s}=0.001<<\alpha_1,\alpha_2$.
(b) $\alpha_{t}=0.055$ close to $\alpha_1$. Corresponding $\alpha_{s}=0.03$.
(c) $\alpha_{t}=0.064=\alpha_1$. Corresponding $\alpha_{s}=0.04$.
(d) $\alpha_{t} = 0.137=\alpha_2$. Corresponding $\alpha_{s} = 0.1$.
(e) $\alpha_{t}= \alpha_{s}=0.5>>\alpha_1,\alpha_2$.
(f)  The number of peaks v.s. $\alpha$ (red stars). The vertical lines indicate the critical $\alpha$ values of peak disappearance from simulation (blue solid line) and theoretical prediction (red dashed line). The left blue solid line represents the disappearing of the first peak and the right blue solid line represents the disappearing of the second peak, leaving the distribution a single peak distribution.
(g) The simulated (stars with dashed line) and theoretically predicted (solid line) dependence of the critical $\alpha$ value on the total substrate number $N_{Xt}$. Numbers of all other species and the volume are increased proportionally to keep all the concentrations constant.
}
     \label{Figure2}
 \end{center}
\end{figure}

Here we provide quantitative analysis of the above time scale argument. Let's define
\begin{equation}
	g_1=-\frac{k_+[E_t]x}{K_{M+}+x} , h_1 =  \frac{k_-[A_t]([X]_t-x}{K_{M-}+[X_t]-x},
\end{equation}
and similar expression for the case that the kinase assumes conformer 2 except $k_+$ is replaced by $k_+'$.
 Then the kinetics of a PdPC with one two-state kinase molecule is governed by a set of Liouville equations under the Langevin dynamics approximation \cite{Zwanzig1990, Kepler2001, Gillespie2000},
\begin{eqnarray}
	\partial_{t} \rho_1(x) &=& \frac{1}{2\Omega}\partial_{xx} \left[ (g_1+h_1)\rho_1(x)\right] \nonumber\\ &&-\partial_{x}\left[ (-g_1+h_1) \rho_1(x) \right]  - \alpha \rho_1(x) + \beta \rho_2(x),\label{up}\\
	\partial_{t} \rho_2(x) &=& \frac{1}{2\Omega}\partial_{xx} \left[ (g_2+h_2)\rho_2(x)\right]\nonumber\\ &&-\partial_{x}\left[(-g_2+h_2) \rho_2(x) \right]  + \alpha \rho_1(x) - \beta \rho_2(x),\label{down}
\end{eqnarray}
where $\rho_i(x)$ is the probability density to find the system at kinase confomer $i$ and the number of substrate form $X$ being $x$, $\Omega$ is the system volume. For mathematical simplicity in the following derivations, we assume that for a given kinase conformation, the substrate dynamics can be described continuously and deterministically. This approximation is partially justified by the relative large number of substrates. Then one can drop the diffusion term containing $\Omega$, and solve the above equations analytically (see Appendix C). The theoretical steady state solutions of $\rho(x)=\rho_1(x)+\rho_2(x)$ are also plotted in  \ref{Figure2}. Since in our analysis we neglected
stochasticity of enzyme reactions due to the finite number of substrates, the analytical solutions are bound by the two roots $(x_1,x_2)$ of the equations,
\begin{align}
f_1(x_1) =0, f_2(x_2) =0 \label{u-d-point}.
\end{align}
In the case of fast switching rates, the solutions vanish at the turning points $(x_1,x_2)$, and become identically zeros outside of the interval $[x_{1},x_{2}]$. Physically it means that the enzyme equilibrates quickly and the rest of the system `feels' only averaged reactivity of two enzyme's states.

In the regime of slow switching rates, the steady state solutions $\rho_1(x)$ and $\rho_1(x)$ have two integratable singular points at $(x_1,x_2)$. The solutions diverge at these points (although integration of $\rho_i$ over $x$ is still finite). Of course, the neglected `diffusion' term due to substrate fluctuations becomes important in this situation. This term smears the singularities at the turning points $(x_1,x_2)$. Kepler $\it{et\hspace{0.1cm} al.}$ noticed similar behaviors in their simulation results. However, in order to estimate the critical values of the switching rates $\alpha_{c}$ and $\beta_{c}$, which correspond to transition between unimodal and bimodal distribution, the set of equations \eq{up} and \eq{down} are sufficient.

Even within this approximation the analytical prediction of the transition points agrees well with the simulation results. The conditions for disappearance of these two single points are,
\begin{align}
	\alpha \ge \alpha_c=\partial_{x}f_1(x_1), \beta \ge \beta_{c}=\partial_{x}f_2(x_2) \label{f_critical}.
\end{align}
Note that $1/ \alpha$  (or $1/ \beta$) is the average dwelling time of an enzyme configuration $\tau_K$, and  $1/\alpha_c$ (or $1/\beta_c$) is the relaxation time after the system linearly deviates from the single conformer steady state (at $x_1$ or $x_2$), which are $\tau_S$ in the previous discussions.
 \ref{Figure2}f \& g show good agreement between the critical points obtained by simulation and by \eq{f_critical}. The agreement becomes better for larger number of substrates, suggesting that the discrepancy between the simulated and theoretical results are due to neglecting substrate fluctuations in the theoretical treatment.

\begin{figure}
 \begin{center}
     \includegraphics*[width=3.25in]{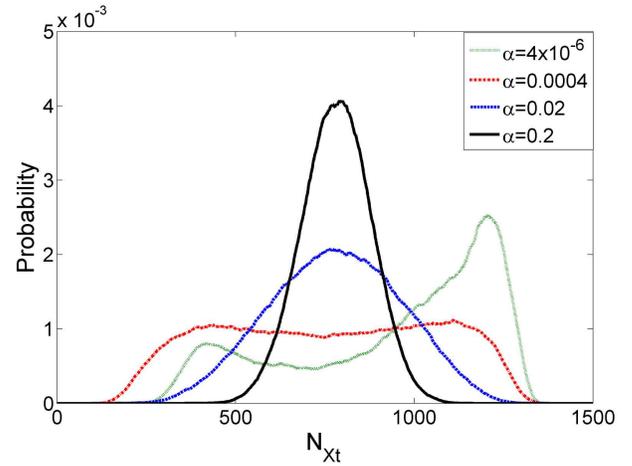}
     \caption{A PdPC with 50 two-state kinase (and phosphatase) enzymes and 1500 substrates. Other parameters are the same as in 1-enzyme case. the $\alpha$ values are: dotted line $4 \times 10^{-6}$, dash-dot line $0.0004$, dashed line $0.02$, and solid line $0.2$.
}
     \label{Nenzyme}
 \end{center}
\end{figure}

\begin{figure}
 \begin{center}
     \includegraphics*[width=3.25in]{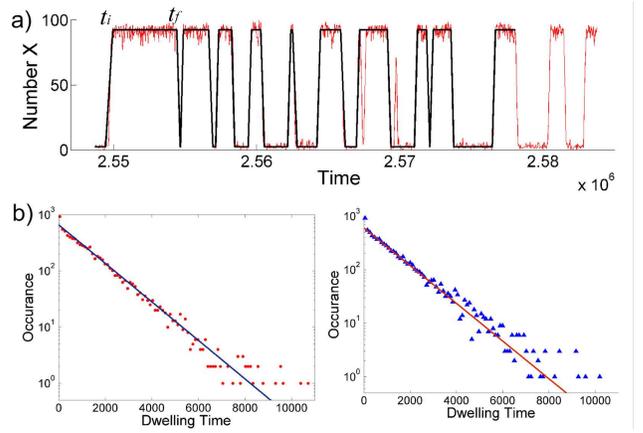}
     \caption{Dwelling time distribution of a PdPC with a single two-state kinase enzyme ($\alpha=0.001$).
     (a) A typical trajectory with steps indicated (dark solid line). The initial and final times of one dwelling state are indicated as $t_i$ and $t_f$, respectively.
     (b) The dwelling time distribution and exponential fitting of the upper (left panel) and lower (right panel) states. The fitting slopes are $-7.9\times 10^{-4}$ and $-8.1\times 10^{-4}$, respectively.
}
     \label{Dwell}
 \end{center}
\end{figure}	
\begin{figure}
 \begin{center}
     \includegraphics*[width=3.25in]{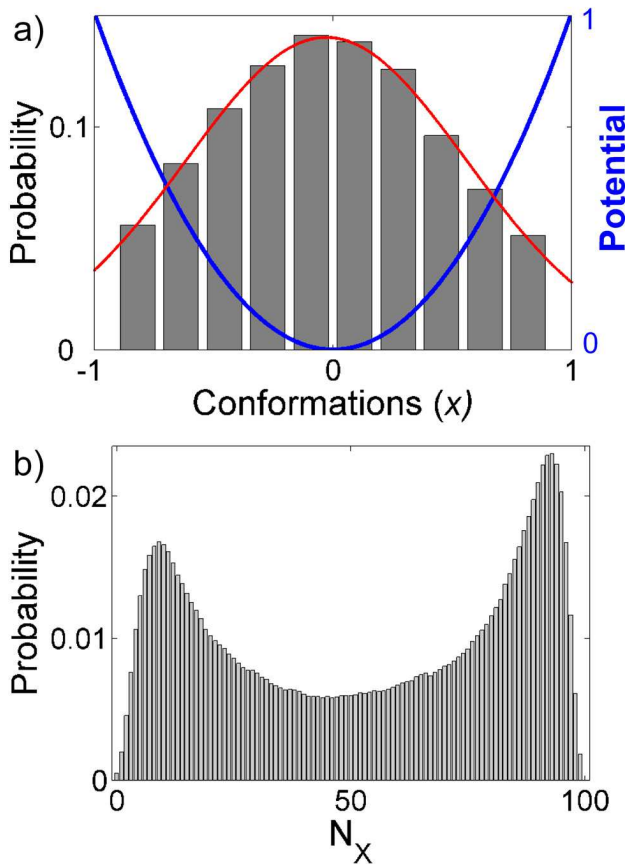}
     \caption{Multi-State Enzyme fluctuation produced bimodal distribution.
     (a) Enzyme conformation fluctuation along a harmonic potential (blue line) and the Boltzmann distribution (gray bars and red line).
     (b) The corresponding bimodal distribution of reactant $X$. }
     \label{figure:Multi-state}
 \end{center}
\end{figure}	
%

\begin{table}[ht]
 \centering
 \begin{tabular}{c c c}
 \hline\hline
    Parameters & Values &  \\ [0.5ex]
    \hline
    rate constant& in reduced unit & \\
    $k_{1f}$ & 50 &   \\
    $k_{+f}$ & 1.6 &   \\
    $k_{1f'}$ & 50 &   \\
    $k_{+f'}$ & 0.4 &   \\
    $k_{2f}$ & 50 &   \\
    $k_{-f}$ & 1 &   \\
    & & \\
    Free energy & in $k_BT$ & \\
     $\Delta\mu_{ATP}$ & $-20 $ & \\
    $\mu_{E1\_X}$  & $0 $ &   \\
    $\mu_{E1X}$   & $-5 $ &   \\
    $\mu_{E1\_X*}$   &  $0 $ &   \\
    $\mu_{A\_X*}$   &  $0 $ &   \\
    $\mu_{AX*}$   &  $-5 $ &   \\
    $\mu_{A\_X}$  &  $0 $ &   \\
    $\mu_{E2\_X}$   &  $0 $ &   \\
    $\mu_{E2X}$   &  $-5 $ &   \\
    [1ex]
    \hline
    * See Appendix A
 \end{tabular}
  \caption{Simulation parameters}
\end{table}

\subsection{Multi-enzyme systems}
For $N$ independent two state kinases the probability to have $k$ kinases in conformer 1 has following binomial distribution
\begin{align}
	\pi(k,t) = \binom{k}{N}p_1^{k}(1-p_1)^{N-k},\label{binomial}
\end{align}
where we defined
\begin{align}
	p_{1}(t) = p^{s}_{1} - (p^{s}_{1} - p^{i}_{1})e^{-(\alpha+\beta)t},\label{exp}
\end{align}
where $p^{i}_{1}$ is initial probability to be in the state $1$, and the steady state probability $p^{s}_{1}$ is given by $p_1=\beta/(\alpha+\beta)$.

Note that binomial distribution \eq{binomial} becomes normal in the case $N\rightarrow\infty$.
Therefore, in this limit one can represent overall concentration of the enzymes as (cf. with equation {\bf S8} in \cite{Samoilov2005})
\begin{align}
	E_{t} = E + \xi(t),
\end{align}
where $\xi(t)$ is Gaussian noise with correlation function decaying \emph{exponentially} fast, see \eq{exp}.
Hence, only when the switching rates $\alpha$ and $\beta$ are fast ($\it{i.e.}$, no dynamic disorder) the white noise approximation used in reference \cite{Samoilov2005} is expected to work well, since in this case exponential decay of the correlation function can be safely replaced by $\delta$-function. Recently Warmflash $\it{et. al}$ also discussed the legitimacy of using the $\delta$-function approximation \cite{Warmflash2008}.

Let us calculate noise-noise correlator explicitly. One gets
\begin{align}
	\ave{n(t)n(t')} \equiv \sum_{n,n'}n n' P(n,t | n',t')P(n',t'),
\end{align}
where $P(n,t | n',t')$ is the conditional probability to have $n$ enzymes in conformer 1 at time $t$, provided that
at time $t'$ the number of enzymes in conformer 1 is {\em exactly} $n'$. The second term in the product,
$P(n',t')$, is the probability to have $n'$ enzymes in conformer 1 at time $t'$ and it depends on initial conditions.
However, for late times $t>t'\gg (\alpha+\beta)^{-1}$ it can be safely replaced by the steady state distribution:
\begin{align}
	P(n',t') \sim \pi_{s}(n').
\end{align}
As for conditional probability, we derive
\begin{align}
	P(n,t | n',t') = \pi(n,t-t'),
\end{align}
where dependence on the number $n'$ comes from the initial conditions in \eq{exp}:
\begin{align}
	p^{i}_{1} =  2^{-N}\binom{n'}{N}.
\end{align}
It means that $p_{1}(t-t')$ in \eq{exp} depends implicitly on $n'$, and hence will be defined as $p_{1}(n',t-t')$.
After some algebra we get
\begin{align}
	\ave{n(t)n(t')} = N\sum_{n'}n' p_{1}(n',t-t') \binom{n'}{N}(p_1^{s})^{n'}(1-p_1^{s})^{N-n'}.\label{corr}
\end{align}
It is easy to check that the first term in the expression for $p_{1}(n',t-t')$, namely  $p_{1}^{s}$, adds the
contribution $\ave{n(t)}\ave{n(t')}$ for a late times. Therefore, one obtains a two time correlation function
\begin{align}
	\ave{n(t)n(t')} - \ave{n(t)}\ave{n(t')} = \sigma^{2} e^{-(\alpha+\beta)(t-t')},
\end{align}
exponentially decaying in time with correlation strength $\sigma$ that can be explicitly obtained from the \eq{corr}.

\section{Discussion and concluding remarks}

Slow conformational fluctuations have been suggested to be general properties of proteins, and result in dynamic disorder. However, so far only metabolic enzymes have been directly examined at the single molecule level. If demonstrated, the existence of dynamic disorder in general may greatly modify our understanding of dynamics of biological networks (e.g., signal transduction networks). It provides a new source of in general non-white noises. In this work, we exploit the ultrasensitivity of a PdPC (or a similar system as discussed previously) to amplify molecular level slow conformational fluctuations. The method may be used experimentally for quick screening and qualitative/semi-quantitative estimation of molecular fluctuations in signal transduction networks. Here we propose a possible experimental setup. One adds one or a few kinases, corresponding phosphatase (with its amount adjusted so the average activity ratio $\overline{\theta}\sim 1$),  a relatively large amount of substrate molecules, and ATP regeneration system in an isolated chamber.
Experimentally one may consider the micro-fabrication technique  produced high density small reaction chambers used previously in single molecule protein motor and enzyme studies \cite{Rondelez2005, Rissin2008}. Containers may stochastically contain different number of molecules of the kinase, phosphatase, with some of them giving the desired $\overline{\theta}\sim 1$.
Monitoring substrate fluctuations ($e.g.$, through fluorescence) may reveal information about molecular level fluctuations.  In general protein fluctuation is more complicated than the two-state model used here. The latter should be viewed as coarse-grained model. In this work for simplicity we didn't consider possible conformational fluctuations of the phosphatase and even the substrate (which may act as enzymes for other reactions). Including these possibilities make the analysis more difficult, but won't change the conclusion that molecular level fluctuations can couple to fluctuations at network levels and be amplified by the latter.

There are several studies on systems showing stochasticity induced bimodal distribution without deterministic counterpart. \cite{Samoilov2005, Kepler2001, Paulsson2005, Blake2003, Artyomov2007, Karmakar2006, Qian2009}  In eukaryotic transcription, a gene may be turned on and off through binding and dissociation of a regulating protein, which may result in bimodal distribution of the expressed protein level. The process is mathematically equivalent to the  problem  we discussed here.  Physically the mechanism of generating a bimodal distribution is trivial. The system (PdPC) has a fluctuating parameter, the ratio of the overall enzyme activity $\theta$ (not $\overline{\theta}$). When the parameter fluctuates sufficiently slow, the distribution is approximately a mixture of localized distributions corresponding to different parameter values, and thus may have more than one peak. This situation is fundamentally different from macroscopic bistable systems, which have more than one steady state for a given set of parameters, and usually some feedback mechanism is involved. Possible biological significances of a network generating bimodal distributions without deterministic counterpart has been suggested in the literature \cite{Samoilov2005, Paulsson2005, Artyomov2007}. It remains to be examined whether the mechanism discussed in this work is biologically relevant, or reversely evolution has selected signal transduction proteins showing minimal dynamic disorder \cite{Xing2008}. As shown in   \ref{Figure2}g and   \ref{Nenzyme}, with fixed enzyme molecular property, a system reduces to unimodal distributions on increasing the system size. Therefore the mechanism of dynamic disorder induced bimodal distribution plays a significant role only for small sized systems. We want to point out that Morishita ${\it et. al.}$ \cite{Morishita2006} has theoretically suggested that signal transduction cascades have optimal performance with only $\sim 50$ copies per specie, which makes the dynamic disorder mechanism plausible. In a real system it is more likely that noises arising from dynamic disorder, which has broad time-scale distribution,  will couple with sources from other processes, such as enzyme synthesis and degradation, and may result in complex dynamic behaviors. Therefore physical chemistry studies of molecular level protein dynamics may provide important and necessary information for understanding cellular level dynamics.

\appendix

\section{Appendixes}
\subsection*{A}
Here we show how we make connections between the equations for bulk analysis  and the ones for stochastic simulations with molecular number. For example, if we have a reaction
\begin{equation*}
    X+E_1\reactionrates{k^0_{1f}}{k^0_{1b}} XE_1,
\end{equation*}
we can write down the ODE equations for this reaction
\begin{equation*}
    \frac{d[XE_1]}{dt}=k^0_{1f} [X] [E_1]-k^0_{1b} [XE_1].
\end{equation*}
First of all, we choose $1/k_{-f}$ as our time unit, where  $k_{-f}$ is the rate constant for the backward enzymatic step. Then,
\begin{equation*}
    \frac{d[XE_1]}{dt}=k_{1f} [X] [E_1]-k_{1b} [XE_1].
\end{equation*}
with $k_{1f} = k^0_{1f}/k_{-f}$,   $k_{1b} = k^0_{1b}/k_{-f}$.
If we want to deal with variables in the unit of molecular numbers instead of concentration, we then have
\begin{equation*}
    \frac{d\left( \frac{N_{XE_1}}{N_A V_{0}} \right)}{dt}=k_{1f} \frac{N_{X}}{N_A V_{0}} \frac{N_{E_1}}{N_A V_{0}}-k_{1b} \frac{N_{XE_1}}{N_A V_{0}}
\end{equation*}
where $N_A$ is Avogadro constant. $V_0$ is the volume of the system. We can further simplify the expression,
\begin{equation*}
    \frac{dN_{XE_1}}{dt} = k_{1f} \frac{N_{X} N_{E_1}}{N_A V_{0}}- k_{1b} N_{XE_1}
\end{equation*}
In all of our simulations, we kept a constant value for the substrate concentration $N_{X_t}/(N_A V_0)=1$. Then,
\begin{equation*}
    \frac{dN_{XE_1}}{dt} = \frac{k_{1f}}{N_{X_t}}  N_{X} N_{E_1}- k_{1b} N_{XE_1}
\end{equation*}

\subsection*{B: Multi-state enzyme fluctuation}
In general, an enzyme fluctuates continuously along conformational coordinates. One should consider the two-state model discussed in the main text as a coarse-grained model. Here we will use a more complicated model to show that our main conclusions still hold in general case. We consider an enzyme diffuse slowly along a harmonic potential of coordinate $x$, $G(x)=x^2$, where we have chose the units so $G=1 k_B T$ at $x=1$. Motion along the conformational coordinate couples to the enzymatic reaction rate constants with an exponential factor $k_{+f} = k^0_{+f} \exp(\lambda x)$, where $\lambda = -0.5$. One specific example is that $x$ is the donor-acceptor distance for an electron-transfer reaction. We model the diffusion as hopping among 10 discrete states. The conversion rate constant between enzyme conformations $\alpha_i$ is at $10^{-2}$. The backward rate constant is then determined by the detailed balance requirement $\beta_i = \alpha_i \exp (- (x_i^2 - x_{i-1}^2)/k_B T)$. Other parameters are the same as in the 2-state enzyme simulations. The reactions are listed below,
\begin{eqnarray*}
& & X+E_1\reactionrates{k_{1f}}{k_{1b}} XE_1\reactionrates{k_{+1f}}{k_{+1b}}   X^* + E_1,\\
& & X+E_2\reactionrates{k_{2f}}{k_{2b}}  XE_2\reactionrates{k_{+2f}}{k_{+2b}}   X^* + E_2,\\
& & ... ... \\
& & X+E_{m-1}\reactionrates{k_{(m-1)f}}{k_{(m-1)b}} XE_{m-1}\reactionrates{k_{+(m-1)f}}{k_{+(m-1)b}}   X^* + E_{m-1},\\
& & X+E_{m}\reactionrates{k_{mf}}{k_{mb}}  XE_m\reactionrates{k_{+mf}}{k_{+mb}}   X^* + E_m,\\
& & X^*+A\reactionrates{k_{af}}{k_{ab}} X^*A \reactionrates{k_{-f}}{k_{-b}}  X + A, \\
& & E_1\reactionrates{\alpha_1}{\beta_1} E_2 \reactionrates{\alpha_2}{\beta_2} ...... \reactionrates{\alpha_{m-2}}{\beta_{m-2}} E_{m-1}  \reactionrates{\alpha_{m-1}}{\beta_{m-1}} E_m, \\
& & E_1X\reactionrates{\alpha_1'}{\beta_1'} E_2X \reactionrates{\alpha_2'}{\beta_2'} ...... \reactionrates{\alpha_{m-2}'}{\beta_{m-2}'} E_{m-1}X  \reactionrates{\alpha_{m-1}'}{\beta_{m-1}'} E_mX.
\end{eqnarray*}

\ref{figure:Multi-state} shows that the reactant $X$ shows a bimodal distribution if the enzymatic conformational fluctuation is slow. This result reiterates our suggestion that the ultrasensitive network amplifies small enzymatic activity fluctuations into large substrate number fluctuations.

\subsection*{C}
  By omitting the diffusion terms in the \eq{up} and \eq{down} one derives for steady state:
\begin{align}
	\partial_{x}\left[f_{1}(x) \rho_{1} \right] = \alpha \rho_{1} - \beta \rho_{2},\\
	f_{1}(x) \rho_{1} + f_{2}(x) \rho_{2} = \mathrm{const}.\label{const}
\end{align}
For very fast switching rates $\alpha$ and $\beta$ we expect an unimodal distribution centered somewhere in between two `turning' points $x_1$ and $x_2$, see \eq{u-d-point}. Therefore, the steady state solution should vanish at these points and be identically zero outside an interval $[x_2,x_1]$.

In order to satisfy these boundary conditions, one has to set a constant in \eq{const} to be zero. Hence, we obtain
\begin{equation}
	\partial_{x}\left[f_{1}(x) \rho_{1} \right] = \left[\alpha +\beta\frac{f_{1}(x)}{f_{2}(x)} \right]\rho_{1}.\label{uni}
\end{equation}
The solution of the \eq{uni} depends, of course, on particular choice of the function $f(x)$. However, it is guaranteed that there is an {\em unique} root of the function $f(x)$ in the corresponding physical region of variable $x$ \cite{Samoilov2005}. Hence, the differential equation \eq{uni} is singular  only at two points $x_2$ and
$x_1$, which are the boundary points. One can find an asymptotic behavior of the steady state solution near these points:
\begin{equation}
	\rho_{1} \sim (x-x_2)^{a_2}(x_1-x)^{a_1} g(x),
\end{equation}
where $g(x)$ is analytic function of $x$ in the interval $[x_2,x_1]$, satisfying condition $g(x_2) = g(x_1) = 1$. The exponents $a_2$ and $a_1$ are
\begin{align}
	a_2 = \frac{\beta}{f_2(x_{2})} - 1,\\
	a_1 = \frac{\alpha}{f_1(x_{1})} - 1.
\end{align}

Therefore, if the conditions \eq{f_critical} are satisfied, one expects unimodal distribution. Otherwise, there exists at least one additional peak in the distribution. In this case of the slow switching rates the equation \eq{uni} predicts divergence of the solution at one or both boundary points $x_2$ and $x_1$. This is an indication that diffusion terms that we omitted for our estimate become relevant. The diffusion terms makes the overall distribution finite.

\begin{acknowledgements}
We thank Alex Elgart, Jian Liu, and Wei Min for useful discussions.
\end{acknowledgements}


\bibliography{Bimodal_JPCB_v2_1}

\begin{thebibliography}{46}
\expandafter\ifx\csname natexlab\endcsname\relax\def\natexlab#1{#1}\fi
\expandafter\ifx\csname bibnamefont\endcsname\relax
  \def\bibnamefont#1{#1}\fi
\expandafter\ifx\csname bibfnamefont\endcsname\relax
  \def\bibfnamefont#1{#1}\fi
\expandafter\ifx\csname citenamefont\endcsname\relax
  \def\citenamefont#1{#1}\fi
\expandafter\ifx\csname url\endcsname\relax
  \def\url#1{\texttt{#1}}\fi
\expandafter\ifx\csname urlprefix\endcsname\relax\def\urlprefix{URL }\fi
\providecommand{\bibinfo}[2]{#2}
\providecommand{\eprint}[2][]{\url{#2}}

\bibitem[{\citenamefont{Lu et~al.}(1998)\citenamefont{Lu, Xun, and
  Xie}}]{LuXun1998}
\bibinfo{author}{\bibfnamefont{H.~P.} \bibnamefont{Lu}},
  \bibinfo{author}{\bibfnamefont{L.}~\bibnamefont{Xun}}, \bibnamefont{and}
  \bibinfo{author}{\bibfnamefont{X.~S.} \bibnamefont{Xie}},
  \bibinfo{journal}{Science} \textbf{\bibinfo{volume}{282}},
  \bibinfo{pages}{1877} (\bibinfo{year}{1998}).

\bibitem[{\citenamefont{Chen et~al.}(2003)\citenamefont{Chen, Hu, Vorpagel, and
  Lu}}]{Chen2003}
\bibinfo{author}{\bibfnamefont{Y.}~\bibnamefont{Chen}},
  \bibinfo{author}{\bibfnamefont{D.}~\bibnamefont{Hu}},
  \bibinfo{author}{\bibfnamefont{E.~R.} \bibnamefont{Vorpagel}},
  \bibnamefont{and} \bibinfo{author}{\bibfnamefont{H.~P.} \bibnamefont{Lu}},
  \bibinfo{journal}{The Journal of Physical Chemistry B}
  \textbf{\bibinfo{volume}{107}}, \bibinfo{pages}{7947} (\bibinfo{year}{2003}).

\bibitem[{\citenamefont{Tan et~al.}(2003)\citenamefont{Tan, Nalbant,
  Toutchkine, Hu, Vorpagel, Hahn, and Lu}}]{Tan2003}
\bibinfo{author}{\bibfnamefont{X.}~\bibnamefont{Tan}},
  \bibinfo{author}{\bibfnamefont{P.}~\bibnamefont{Nalbant}},
  \bibinfo{author}{\bibfnamefont{A.}~\bibnamefont{Toutchkine}},
  \bibinfo{author}{\bibfnamefont{D.}~\bibnamefont{Hu}},
  \bibinfo{author}{\bibfnamefont{E.~R.} \bibnamefont{Vorpagel}},
  \bibinfo{author}{\bibfnamefont{K.~M.} \bibnamefont{Hahn}}, \bibnamefont{and}
  \bibinfo{author}{\bibfnamefont{H.~P.} \bibnamefont{Lu}},
  \bibinfo{journal}{The Journal of Physical Chemistry B}
  \textbf{\bibinfo{volume}{108}}, \bibinfo{pages}{737} (\bibinfo{year}{2003}).

\bibitem[{\citenamefont{Lu}(2005)}]{Lu2005}
\bibinfo{author}{\bibfnamefont{H.~P.} \bibnamefont{Lu}}, \bibinfo{journal}{Acc
  Chem Res} \textbf{\bibinfo{volume}{38}}, \bibinfo{pages}{557}
  (\bibinfo{year}{2005}).

\bibitem[{\citenamefont{Liu et~al.}(2006)\citenamefont{Liu, Hu, Tan, and
  Lu}}]{Liu2006}
\bibinfo{author}{\bibfnamefont{R.}~\bibnamefont{Liu}},
  \bibinfo{author}{\bibfnamefont{D.}~\bibnamefont{Hu}},
  \bibinfo{author}{\bibfnamefont{X.}~\bibnamefont{Tan}}, \bibnamefont{and}
  \bibinfo{author}{\bibfnamefont{H.~P.} \bibnamefont{Lu}}, \bibinfo{journal}{J
  Am Chem Soc} \textbf{\bibinfo{volume}{128}}, \bibinfo{pages}{10034}
  (\bibinfo{year}{2006}).

\bibitem[{\citenamefont{English et~al.}(2006)\citenamefont{English, Min, van
  Oijen, Lee, Luo, Sun, Cherayil, Kou, and Xie}}]{English2006}
\bibinfo{author}{\bibfnamefont{B.~P.} \bibnamefont{English}},
  \bibinfo{author}{\bibfnamefont{W.}~\bibnamefont{Min}},
  \bibinfo{author}{\bibfnamefont{A.~M.} \bibnamefont{van Oijen}},
  \bibinfo{author}{\bibfnamefont{K.~T.} \bibnamefont{Lee}},
  \bibinfo{author}{\bibfnamefont{G.~B.} \bibnamefont{Luo}},
  \bibinfo{author}{\bibfnamefont{H.~Y.} \bibnamefont{Sun}},
  \bibinfo{author}{\bibfnamefont{B.~J.} \bibnamefont{Cherayil}},
  \bibinfo{author}{\bibfnamefont{S.~C.} \bibnamefont{Kou}}, \bibnamefont{and}
  \bibinfo{author}{\bibfnamefont{X.~S.} \bibnamefont{Xie}},
  \bibinfo{journal}{Nat. Chem. Biol.} \textbf{\bibinfo{volume}{2}},
  \bibinfo{pages}{87} (\bibinfo{year}{2006}).

\bibitem[{\citenamefont{Rissin et~al.}(2008)\citenamefont{Rissin, Gorris, and
  Walt}}]{Rissin2008}
\bibinfo{author}{\bibfnamefont{D.~M.} \bibnamefont{Rissin}},
  \bibinfo{author}{\bibfnamefont{H.~H.} \bibnamefont{Gorris}},
  \bibnamefont{and} \bibinfo{author}{\bibfnamefont{D.~R.} \bibnamefont{Walt}},
  \bibinfo{journal}{J Am Chem Soc} \textbf{\bibinfo{volume}{130}},
  \bibinfo{pages}{5349} (\bibinfo{year}{2008}).

\bibitem[{\citenamefont{Wang and Lu}(2008)}]{Wang2008}
\bibinfo{author}{\bibfnamefont{X.}~\bibnamefont{Wang}} \bibnamefont{and}
  \bibinfo{author}{\bibfnamefont{H.~P.} \bibnamefont{Lu}}, \bibinfo{journal}{J
  Phys Chem B} \textbf{\bibinfo{volume}{112}}, \bibinfo{pages}{14920}
  (\bibinfo{year}{2008}).

\bibitem[{\citenamefont{Chen et~al.}(2009)\citenamefont{Chen, Xu, Zhou, Panda,
  and Kalininskiy}}]{chen2009}
\bibinfo{author}{\bibfnamefont{P.}~\bibnamefont{Chen}},
  \bibinfo{author}{\bibfnamefont{W.}~\bibnamefont{Xu}},
  \bibinfo{author}{\bibfnamefont{X.}~\bibnamefont{Zhou}},
  \bibinfo{author}{\bibfnamefont{D.}~\bibnamefont{Panda}}, \bibnamefont{and}
  \bibinfo{author}{\bibfnamefont{A.}~\bibnamefont{Kalininskiy}},
  \bibinfo{journal}{Chemical Physics Letters} \textbf{\bibinfo{volume}{470}},
  \bibinfo{pages}{151} (\bibinfo{year}{2009}).

\bibitem[{\citenamefont{Chen and Andoy}(2008)}]{Chen2008}
\bibinfo{author}{\bibfnamefont{P.}~\bibnamefont{Chen}} \bibnamefont{and}
  \bibinfo{author}{\bibfnamefont{N.~M.} \bibnamefont{Andoy}},
  \bibinfo{journal}{Inorganica Chimica Acta} \textbf{\bibinfo{volume}{361}},
  \bibinfo{pages}{809} (\bibinfo{year}{2008}).

\bibitem[{\citenamefont{Roeffaers et~al.}(2007)\citenamefont{Roeffaers,
  De~Cremer, Uji-i, Muls, Sels, Jacobs, De~Schryver, De~Vos, and
  Hofkens}}]{Roeffaers2007}
\bibinfo{author}{\bibfnamefont{M.~B.~J.} \bibnamefont{Roeffaers}},
  \bibinfo{author}{\bibfnamefont{G.}~\bibnamefont{De~Cremer}},
  \bibinfo{author}{\bibfnamefont{H.}~\bibnamefont{Uji-i}},
  \bibinfo{author}{\bibfnamefont{B.}~\bibnamefont{Muls}},
  \bibinfo{author}{\bibfnamefont{B.~F.} \bibnamefont{Sels}},
  \bibinfo{author}{\bibfnamefont{P.~A.} \bibnamefont{Jacobs}},
  \bibinfo{author}{\bibfnamefont{F.~C.} \bibnamefont{De~Schryver}},
  \bibinfo{author}{\bibfnamefont{D.~E.} \bibnamefont{De~Vos}},
  \bibnamefont{and} \bibinfo{author}{\bibfnamefont{J.}~\bibnamefont{Hofkens}},
  \bibinfo{journal}{Proceedings of the National Academy of Sciences}
  \textbf{\bibinfo{volume}{104}}, \bibinfo{pages}{12603}
  (\bibinfo{year}{2007}).

\bibitem[{\citenamefont{Xu et~al.}(2009)\citenamefont{Xu, Kong, and
  Chen}}]{Xu2009}
\bibinfo{author}{\bibfnamefont{W.}~\bibnamefont{Xu}},
  \bibinfo{author}{\bibfnamefont{J.~S.} \bibnamefont{Kong}}, \bibnamefont{and}
  \bibinfo{author}{\bibfnamefont{P.}~\bibnamefont{Chen}},
  \bibinfo{journal}{Phys Chem Chem Phys} \textbf{\bibinfo{volume}{11}},
  \bibinfo{pages}{2767} (\bibinfo{year}{2009}).

\bibitem[{\citenamefont{Min et~al.}(2005)\citenamefont{Min, Luo, Cherayil, Kou,
  and Xie}}]{MinLuo2005}
\bibinfo{author}{\bibfnamefont{W.}~\bibnamefont{Min}},
  \bibinfo{author}{\bibfnamefont{G.~B.} \bibnamefont{Luo}},
  \bibinfo{author}{\bibfnamefont{B.~J.} \bibnamefont{Cherayil}},
  \bibinfo{author}{\bibfnamefont{S.~C.} \bibnamefont{Kou}}, \bibnamefont{and}
  \bibinfo{author}{\bibfnamefont{X.~S.} \bibnamefont{Xie}},
  \bibinfo{journal}{Phys. Rev. Lett.} \textbf{\bibinfo{volume}{94}},
  \bibinfo{pages}{198302} (\bibinfo{year}{2005}).

\bibitem[{\citenamefont{Xing and Kim}(2006)}]{XingKim2006}
\bibinfo{author}{\bibfnamefont{J.}~\bibnamefont{Xing}} \bibnamefont{and}
  \bibinfo{author}{\bibfnamefont{K.~S.} \bibnamefont{Kim}},
  \bibinfo{journal}{Phys. Rev. E} \textbf{\bibinfo{volume}{74}},
  \bibinfo{pages}{061911} (\bibinfo{year}{2006}).

\bibitem[{\citenamefont{Frieden}(1979)}]{Frieden1979}
\bibinfo{author}{\bibfnamefont{C.}~\bibnamefont{Frieden}},
  \bibinfo{journal}{Ann. Rev. Biochem.} \textbf{\bibinfo{volume}{48}},
  \bibinfo{pages}{471} (\bibinfo{year}{1979}).

\bibitem[{\citenamefont{Frieden}(1970)}]{Frieden1970}
\bibinfo{author}{\bibfnamefont{C.}~\bibnamefont{Frieden}}, \bibinfo{journal}{J.
  Biol. Chem.} \textbf{\bibinfo{volume}{245}}, \bibinfo{pages}{5788}
  (\bibinfo{year}{1970}).

\bibitem[{\citenamefont{Ricard and Cornish-Bowden}(1987)}]{Ricard1987}
\bibinfo{author}{\bibfnamefont{J.}~\bibnamefont{Ricard}} \bibnamefont{and}
  \bibinfo{author}{\bibfnamefont{A.}~\bibnamefont{Cornish-Bowden}},
  \bibinfo{journal}{Eur. J. Biochem.} \textbf{\bibinfo{volume}{166}},
  \bibinfo{pages}{255} (\bibinfo{year}{1987}).

\bibitem[{\citenamefont{Ainslie et~al.}(1972)\citenamefont{Ainslie, Shill, and
  Neet}}]{Ainslie1972}
\bibinfo{author}{\bibfnamefont{J.}~\bibnamefont{Ainslie},
  \bibfnamefont{G.~Robert}}, \bibinfo{author}{\bibfnamefont{J.~P.}
  \bibnamefont{Shill}}, \bibnamefont{and} \bibinfo{author}{\bibfnamefont{K.~E.}
  \bibnamefont{Neet}}, \bibinfo{journal}{J. Biol. Chem.}
  \textbf{\bibinfo{volume}{247}}, \bibinfo{pages}{7088} (\bibinfo{year}{1972}).

\bibitem[{\citenamefont{Zwanzig}(1990)}]{Zwanzig1990}
\bibinfo{author}{\bibfnamefont{R.}~\bibnamefont{Zwanzig}},
  \bibinfo{journal}{Acc. Chem. Res.} \textbf{\bibinfo{volume}{23}},
  \bibinfo{pages}{148} (\bibinfo{year}{1990}).

\bibitem[{\citenamefont{Austin et~al.}(1975)\citenamefont{Austin, Beeson,
  Eisenstein, Frauenfelder, and Gunsalus}}]{Austin1975}
\bibinfo{author}{\bibfnamefont{R.~H.} \bibnamefont{Austin}},
  \bibinfo{author}{\bibfnamefont{K.~W.} \bibnamefont{Beeson}},
  \bibinfo{author}{\bibfnamefont{L.}~\bibnamefont{Eisenstein}},
  \bibinfo{author}{\bibfnamefont{H.}~\bibnamefont{Frauenfelder}},
  \bibnamefont{and} \bibinfo{author}{\bibfnamefont{I.~C.}
  \bibnamefont{Gunsalus}}, \bibinfo{journal}{Biochemistry}
  \textbf{\bibinfo{volume}{14}}, \bibinfo{pages}{5355} (\bibinfo{year}{1975}).

\bibitem[{\citenamefont{Monod et~al.}(1965)\citenamefont{Monod, Wyman, and
  Changeux}}]{Monod1965}
\bibinfo{author}{\bibfnamefont{J.}~\bibnamefont{Monod}},
  \bibinfo{author}{\bibfnamefont{J.}~\bibnamefont{Wyman}}, \bibnamefont{and}
  \bibinfo{author}{\bibfnamefont{J.~P.} \bibnamefont{Changeux}},
  \bibinfo{journal}{J. Mol. Biol.} \textbf{\bibinfo{volume}{12}},
  \bibinfo{pages}{88} (\bibinfo{year}{1965}).

\bibitem[{\citenamefont{Kern and Zuiderweg}(2003)}]{Kern2003}
\bibinfo{author}{\bibfnamefont{D.}~\bibnamefont{Kern}} \bibnamefont{and}
  \bibinfo{author}{\bibfnamefont{E.~R.~P.} \bibnamefont{Zuiderweg}},
  \bibinfo{journal}{Curr. Opin. Struc. Biol.} \textbf{\bibinfo{volume}{13}},
  \bibinfo{pages}{748} (\bibinfo{year}{2003}).

\bibitem[{\citenamefont{Volkman et~al.}(2001)\citenamefont{Volkman, Lipson,
  Wemmer, and Kern}}]{Volkman2001}
\bibinfo{author}{\bibfnamefont{B.~F.} \bibnamefont{Volkman}},
  \bibinfo{author}{\bibfnamefont{D.}~\bibnamefont{Lipson}},
  \bibinfo{author}{\bibfnamefont{D.~E.} \bibnamefont{Wemmer}},
  \bibnamefont{and} \bibinfo{author}{\bibfnamefont{D.}~\bibnamefont{Kern}},
  \bibinfo{journal}{Science} \textbf{\bibinfo{volume}{291}},
  \bibinfo{pages}{2429} (\bibinfo{year}{2001}).

\bibitem[{\citenamefont{Xing}(|2007|)}]{Xing2007}
\bibinfo{author}{\bibfnamefont{J.}~\bibnamefont{Xing}}, \bibinfo{journal}{Phys.
  Rev. Lett.} \textbf{\bibinfo{volume}{99}}, \bibinfo{pages}{168103}
  (\bibinfo{year}{|2007|}).

\bibitem[{\citenamefont{Min et~al.}(|2008|)\citenamefont{Min, Xie, and
  Bagchi}}]{Min2008}
\bibinfo{author}{\bibfnamefont{W.}~\bibnamefont{Min}},
  \bibinfo{author}{\bibfnamefont{X.~S.} \bibnamefont{Xie}}, \bibnamefont{and}
  \bibinfo{author}{\bibfnamefont{B.}~\bibnamefont{Bagchi}},
  \bibinfo{journal}{J. Phys. Chem. B} \textbf{\bibinfo{volume}{112}},
  \bibinfo{pages}{454} (\bibinfo{year}{|2008|}).

\bibitem[{\citenamefont{Goldbeter and Koshland}(1981)}]{Goldbeter1981}
\bibinfo{author}{\bibfnamefont{A.}~\bibnamefont{Goldbeter}} \bibnamefont{and}
  \bibinfo{author}{\bibfnamefont{D.~E.} \bibnamefont{Koshland}},
  \bibinfo{journal}{Proc. Natl. Acad. Sci. U.S.A.}
  \textbf{\bibinfo{volume}{78}}, \bibinfo{pages}{6840} (\bibinfo{year}{1981}).

\bibitem[{\citenamefont{Swain et~al.}(2002)\citenamefont{Swain, Elowitz, and
  Siggia}}]{Swain2002}
\bibinfo{author}{\bibfnamefont{P.~S.} \bibnamefont{Swain}},
  \bibinfo{author}{\bibfnamefont{M.~B.} \bibnamefont{Elowitz}},
  \bibnamefont{and} \bibinfo{author}{\bibfnamefont{E.~D.}
  \bibnamefont{Siggia}}, \bibinfo{journal}{Proc. Natl Acad. Sci. USA}
  \textbf{\bibinfo{volume}{99}}, \bibinfo{pages}{12795} (\bibinfo{year}{2002}).

\bibitem[{\citenamefont{Levine et~al.}(2007)\citenamefont{Levine, Kueh, and
  Mirny}}]{Levine2007}
\bibinfo{author}{\bibfnamefont{J.}~\bibnamefont{Levine}},
  \bibinfo{author}{\bibfnamefont{H.~Y.} \bibnamefont{Kueh}}, \bibnamefont{and}
  \bibinfo{author}{\bibfnamefont{L.}~\bibnamefont{Mirny}},
  \bibinfo{journal}{Biophys. J.} \textbf{\bibinfo{volume}{92}},
  \bibinfo{pages}{4473} (\bibinfo{year}{2007}).

\bibitem[{\citenamefont{Samoilov et~al.}(2005)\citenamefont{Samoilov,
  Plyasunov, and Arkin}}]{Samoilov2005}
\bibinfo{author}{\bibfnamefont{M.}~\bibnamefont{Samoilov}},
  \bibinfo{author}{\bibfnamefont{S.}~\bibnamefont{Plyasunov}},
  \bibnamefont{and} \bibinfo{author}{\bibfnamefont{A.~P.} \bibnamefont{Arkin}},
  \bibinfo{journal}{Proc. Natl. Acad. Sci. U.S.A.}
  \textbf{\bibinfo{volume}{102}}, \bibinfo{pages}{2310} (\bibinfo{year}{2005}).

\bibitem[{\citenamefont{Samoilov and Arkin}(2006)}]{Samoilov2006}
\bibinfo{author}{\bibfnamefont{M.~S.} \bibnamefont{Samoilov}} \bibnamefont{and}
  \bibinfo{author}{\bibfnamefont{A.~P.} \bibnamefont{Arkin}},
  \bibinfo{journal}{Nat Biotech} \textbf{\bibinfo{volume}{24}},
  \bibinfo{pages}{1235} (\bibinfo{year}{2006}).

\bibitem[{\citenamefont{Lepzelter et~al.}(2007)\citenamefont{Lepzelter, Kim,
  and Wang}}]{Lepzelter2007}
\bibinfo{author}{\bibfnamefont{D.}~\bibnamefont{Lepzelter}},
  \bibinfo{author}{\bibfnamefont{K.~Y.} \bibnamefont{Kim}}, \bibnamefont{and}
  \bibinfo{author}{\bibfnamefont{J.}~\bibnamefont{Wang}}, \bibinfo{journal}{J
  Phys Chem B} \textbf{\bibinfo{volume}{111}}, \bibinfo{pages}{10239}
  (\bibinfo{year}{2007}).

\bibitem[{\citenamefont{Thattai and van Oudenaarden}(2001)}]{Thattai2001}
\bibinfo{author}{\bibfnamefont{M.}~\bibnamefont{Thattai}} \bibnamefont{and}
  \bibinfo{author}{\bibfnamefont{A.}~\bibnamefont{van Oudenaarden}},
  \bibinfo{journal}{Proc Natl Acad Sci U S A} \textbf{\bibinfo{volume}{98}},
  \bibinfo{pages}{8614} (\bibinfo{year}{2001}).

\bibitem[{\citenamefont{Barkai and Leibler}(1997)}]{Barkai1997}
\bibinfo{author}{\bibfnamefont{N.}~\bibnamefont{Barkai}} \bibnamefont{and}
  \bibinfo{author}{\bibfnamefont{S.}~\bibnamefont{Leibler}},
  \bibinfo{journal}{Nature} \textbf{\bibinfo{volume}{387}},
  \bibinfo{pages}{913} (\bibinfo{year}{1997}).

\bibitem[{\citenamefont{Miller and Beard}(2008)}]{Miller2008}
\bibinfo{author}{\bibfnamefont{C.~A.} \bibnamefont{Miller}} \bibnamefont{and}
  \bibinfo{author}{\bibfnamefont{D.~A.} \bibnamefont{Beard}},
  \bibinfo{journal}{Biophys J} \textbf{\bibinfo{volume}{95}},
  \bibinfo{pages}{2183} (\bibinfo{year}{2008}).

\bibitem[{\citenamefont{Xing and Chen}(2008)}]{Xing2008}
\bibinfo{author}{\bibfnamefont{J.}~\bibnamefont{Xing}} \bibnamefont{and}
  \bibinfo{author}{\bibfnamefont{J.}~\bibnamefont{Chen}},
  \bibinfo{journal}{PLoS ONE} \textbf{\bibinfo{volume}{3}},
  \bibinfo{pages}{e2140} (\bibinfo{year}{2008}).

\bibitem[{\citenamefont{Gillespie}(1977)}]{Gillespie1977}
\bibinfo{author}{\bibfnamefont{D.~T.} \bibnamefont{Gillespie}},
  \bibinfo{journal}{J. Phys. Chem.} \textbf{\bibinfo{volume}{81}},
  \bibinfo{pages}{2340} (\bibinfo{year}{1977}).

\bibitem[{\citenamefont{Kepler and Elston}(2001)}]{Kepler2001}
\bibinfo{author}{\bibfnamefont{T.~B.} \bibnamefont{Kepler}} \bibnamefont{and}
  \bibinfo{author}{\bibfnamefont{T.~C.} \bibnamefont{Elston}},
  \bibinfo{journal}{Biophys. J.} \textbf{\bibinfo{volume}{81}},
  \bibinfo{pages}{3116} (\bibinfo{year}{2001}).

\bibitem[{\citenamefont{Gillespie}(2000)}]{Gillespie2000}
\bibinfo{author}{\bibfnamefont{D.~T.} \bibnamefont{Gillespie}},
  \bibinfo{journal}{J. Chem. Phys.} \textbf{\bibinfo{volume}{113}},
  \bibinfo{pages}{297} (\bibinfo{year}{2000}).

\bibitem[{\citenamefont{Warmflash et~al.}(2008)\citenamefont{Warmflash,
  Adamson, and Dinner}}]{Warmflash2008}
\bibinfo{author}{\bibfnamefont{A.}~\bibnamefont{Warmflash}},
  \bibinfo{author}{\bibfnamefont{D.~N.} \bibnamefont{Adamson}},
  \bibnamefont{and} \bibinfo{author}{\bibfnamefont{A.~R.}
  \bibnamefont{Dinner}}, \bibinfo{journal}{J Chem Phys}
  \textbf{\bibinfo{volume}{128}}, \bibinfo{pages}{225101}
  (\bibinfo{year}{2008}).

\bibitem[{\citenamefont{Rondelez et~al.}(2005)\citenamefont{Rondelez, Tresset,
  Nakashima, Kato-Yamada, Fujita, Takeuchi, and Noji}}]{Rondelez2005}
\bibinfo{author}{\bibfnamefont{Y.}~\bibnamefont{Rondelez}},
  \bibinfo{author}{\bibfnamefont{G.}~\bibnamefont{Tresset}},
  \bibinfo{author}{\bibfnamefont{T.}~\bibnamefont{Nakashima}},
  \bibinfo{author}{\bibfnamefont{Y.}~\bibnamefont{Kato-Yamada}},
  \bibinfo{author}{\bibfnamefont{H.}~\bibnamefont{Fujita}},
  \bibinfo{author}{\bibfnamefont{S.}~\bibnamefont{Takeuchi}}, \bibnamefont{and}
  \bibinfo{author}{\bibfnamefont{H.}~\bibnamefont{Noji}},
  \bibinfo{journal}{Nature} \textbf{\bibinfo{volume}{433}},
  \bibinfo{pages}{773} (\bibinfo{year}{2005}).

\bibitem[{\citenamefont{Paulsson}(2005)}]{Paulsson2005}
\bibinfo{author}{\bibfnamefont{J.}~\bibnamefont{Paulsson}},
  \bibinfo{journal}{Physics of Life Reviews} \textbf{\bibinfo{volume}{2}},
  \bibinfo{pages}{157} (\bibinfo{year}{2005}).

\bibitem[{\citenamefont{Blake et~al.}(2003)\citenamefont{Blake, KAErn, Cantor,
  and Collins}}]{Blake2003}
\bibinfo{author}{\bibfnamefont{W.~J.} \bibnamefont{Blake}},
  \bibinfo{author}{\bibfnamefont{M.}~\bibnamefont{KAErn}},
  \bibinfo{author}{\bibfnamefont{C.~R.} \bibnamefont{Cantor}},
  \bibnamefont{and} \bibinfo{author}{\bibfnamefont{J.~J.}
  \bibnamefont{Collins}}, \bibinfo{journal}{Nature}
  \textbf{\bibinfo{volume}{422}}, \bibinfo{pages}{633} (\bibinfo{year}{2003}).

\bibitem[{\citenamefont{Artyomov et~al.}(2007)\citenamefont{Artyomov, Das,
  Kardar, and Chakraborty}}]{Artyomov2007}
\bibinfo{author}{\bibfnamefont{M.~N.} \bibnamefont{Artyomov}},
  \bibinfo{author}{\bibfnamefont{J.}~\bibnamefont{Das}},
  \bibinfo{author}{\bibfnamefont{M.}~\bibnamefont{Kardar}}, \bibnamefont{and}
  \bibinfo{author}{\bibfnamefont{A.~K.} \bibnamefont{Chakraborty}},
  \bibinfo{journal}{Proc Natl Acad Sci U S A} \textbf{\bibinfo{volume}{104}},
  \bibinfo{pages}{18958} (\bibinfo{year}{2007}).

\bibitem[{\citenamefont{Karmakar and Bose}(2006)}]{Karmakar2006}
\bibinfo{author}{\bibfnamefont{R.}~\bibnamefont{Karmakar}} \bibnamefont{and}
  \bibinfo{author}{\bibfnamefont{I.}~\bibnamefont{Bose}},
  \bibinfo{journal}{Phys. Biol.} \textbf{\bibinfo{volume}{3}},
  \bibinfo{pages}{200} (\bibinfo{year}{2006}).

\bibitem[{\citenamefont{Qian et~al.}(2009)\citenamefont{Qian, Shi, and
  Xing}}]{Qian2009}
\bibinfo{author}{\bibfnamefont{H.}~\bibnamefont{Qian}},
  \bibinfo{author}{\bibfnamefont{P.-Z.} \bibnamefont{Shi}}, \bibnamefont{and}
  \bibinfo{author}{\bibfnamefont{J.}~\bibnamefont{Xing}},
  \bibinfo{journal}{Physical Chemistry Chemical Physics}
  pp.~\bibinfo{pages}{--} (\bibinfo{year}{2009}),
  \urlprefix\url{http://dx.doi.org/10.1039/b900335p}.

\bibitem[{\citenamefont{Morishita et~al.}(2006)\citenamefont{Morishita,
  Kobayashi, and Aihara}}]{Morishita2006}
\bibinfo{author}{\bibfnamefont{Y.}~\bibnamefont{Morishita}},
  \bibinfo{author}{\bibfnamefont{T.~J.} \bibnamefont{Kobayashi}},
  \bibnamefont{and} \bibinfo{author}{\bibfnamefont{K.}~\bibnamefont{Aihara}},
  \bibinfo{journal}{Biophys J} \textbf{\bibinfo{volume}{91}},
  \bibinfo{pages}{2072} (\bibinfo{year}{2006}).

\end{thebibliography}

\end{document}